\documentclass[aps,prb,twocolumn,superscriptaddress,floatfix,showpacs]{revtex4-1}
\usepackage{graphicx}
\usepackage{amsfonts}
\usepackage{amssymb}
\usepackage{amsmath}
\usepackage{enumerate}
\usepackage{multirow}
\usepackage{color}
\usepackage[all]{xy}
\usepackage{epstopdf}

\newcommand{\be}{\begin{equation}}
\newcommand{\ee}{\end{equation}}
\newcommand{\ba}{\begin{eqnarray}}
\newcommand{\ea}{\end{eqnarray}}
\newcommand{\baa}{\begin{eqnarray*}}
\newcommand{\eaa}{\end{eqnarray*}}

\usepackage{amssymb}
\usepackage{graphicx}
\usepackage{amsmath}
\usepackage{dsfont}
\usepackage{bbm}
\begin{document}
\title{Friedel oscillations at the Dirac-cone-merging point in anisotropic graphene}
\author{Cl\'ement Dutreix}
\affiliation{Laboratoire de Physique des Solides, CNRS UMR-8502,
Universit\'e Paris Sud, 91405 Orsay Cedex, France}
\author{Liviu Bilteanu}
\affiliation{Laboratoire de Physique des Solides, CNRS UMR-8502,
Universit\'e Paris Sud, 91405 Orsay Cedex, France}
\author{Anu Jagannathan}
\affiliation{Laboratoire de Physique des Solides, CNRS UMR-8502,
Universit\'e Paris Sud, 91405 Orsay Cedex, France}
\author{Cristina Bena}
\affiliation{Laboratoire de Physique des Solides, CNRS UMR-8502,
Universit\'e Paris Sud, 91405 Orsay Cedex, France}
\affiliation{Institute de Physique Th\'eorique, CEA/Saclay, Orme des
Merisiers, 91190 Gif-sur-Yvette Cedex, France}

\date{\today}

\begin{abstract}
We study the Friedel oscillations induced by a localized impurity in anisotropic graphene. We focus on the limit when the two inequivalent Dirac points merge. We find that in this limit the Friedel oscillations manifest very peculiar features, such as a strong asymmetry and an atypical inverse square-root decay. Our calculations are performed using both a T-matrix approximation and a tight-binding exact diagonalization technique. They allow us to obtain numerically the local density of states as a function of energy and position, as well as an analytical form of the Friedel oscillations in the continuum limit. The two techniques yield results that are in excellent agreement, confirming the accuracy of such methods to approach this problem.
\end{abstract}
\maketitle

\section{Introduction}
Graphene has known an increased interest over the past years, with some of the most interesting questions at present focusing on the possibility to modify the electronic structure of graphene, either by mechanical deformations such as stretching \cite{2008arXiv0811.4396P}, or twisting \cite{PhysRevLett.99.256802, PhysRevB.84.045436}, via chemical additions, or by changing the nature of substrate. In the light of possible important applications, the most promising directions have been towards opening a gap \cite{2005PhRvL..95v6801K}, enhancing the spin-orbit interaction, the realization of the quantum spin Hall effect \cite{2005PhRvL..95v6801K, 2007PhRvB..75l1403F, 2010PhRvL.104f6805S, 2011PhRvX...1b1001W}, and obtaining integer and fractional quantum Hall states using pseudo-magnetic (curvature) fields \cite{Levy30072010, 2010NatPh...6...30G}. 

One of the most studied modification of graphene is mechanical stretching, which gives rise to a hopping anisotropy, and consequently to a strong renormalization of the band structure. For a critical value of the anisotropy,  such normalization has been predicted to give rise to a hybrid Dirac cone, exhibiting a linear dispersion along one direction, and a quadratic one along the perpendicular one \cite{2008arXiv0811.4396P}. Recently, the realization of a cold-atom equivalent of such anisotropic system has been achieved \cite{2012Natur.483..302T}. Similar hybrid Dirac cones have been predicted to arise when the higher-order hopping parameters are strongly enhanced \cite{PhysRevB.80.153412, PhysRevB.74.033413}, which may occur for example in the presence of adatoms \cite{2012Natur.483..306T}. 

In this work we focus on a system with such hybrid semi-Dirac points and we study the Friedel oscillations (FO) generated in the presence of a single localized impurity. We use both analytical techniques such as the T-matrix approximation, and numerical techniques (the exact diagonalization of the lattice tight-binding Hamiltonian). Using the T-matrix approximation we obtain the form of the Fourier transform of the Friedel oscillations induced by the impurity. We also calculate the real-space form of these oscillations. For small energies and large distances (in the continuum limit) we obtain an exact analytical form of these oscillations,  while we evaluate the short distance behavior of the Friedel oscillations using a numerical integration. On the other hand, we calculate the local density of states (LDOS) at each lattice site using an exact diagonalization of the lattice tight-binding Hamiltonian. Finally we study the form of the LDOS at zero energy using wavefunction arguments along the lines of Ref.~\onlinecite{PhysRevLett.96.036801}, which allow us to obtain an analytical form for the impurity state at zero energy. The results obtained via the above methods are in perfect agreement, confirming the accuracy of these tools for describing the impurity effects in such systems. 

The most interesting characteristic of the observed  Friedel oscillations is a strong anisotropic spatial dependence - the period and decay length of these oscillations depends strongly on direction - consistent with the anisotropy of the band structure. Also we observe an atypical inverse square root  decay for large distances and small energies on each of the two A and B sublattices. Moreover, similar to the isotropic graphene, the LDOS contributions of the two sublattices are dephased by $\pi$, yielding a cancelation of the $1/\sqrt{r}$ terms and an effective $1/r$ decay of these oscillations with the distance from the impurity.

The structure of the paper is as follows. In section II we present the model employed to describe isotropic as well as anisotropic graphene. In Section III we present the Friedel oscillations in the LDOS calculated using wavefunction considerations (A), tight-binding exact diagonalization (B) and the T-matrix approximation (C). We conclude in section IV.

\section{Model}
\begin{figure}[!h]
\centering
\includegraphics[width=6cm]{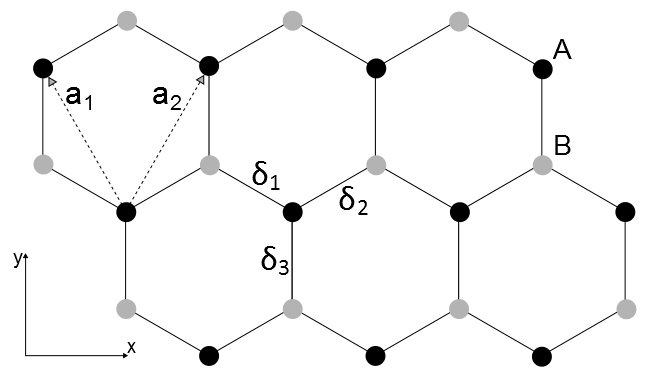}
\caption{Graphene honeycomb lattice.}
\label{Honey}
\end{figure}

Graphene consists of a honeycomb lattice of carbon atoms with two atoms ($A$ and $B$) per unit cell (see Fig.~\ref{Honey}). Denoting the distance between two nearest neighbors $a_{0}$, with $a_0=0.142$ nm, then $\bf a_1$$=a_{0}(-\frac{\sqrt{3}}{2},\frac{3}{2})$ and $\bf a_2$$=a_{0}(\frac{\sqrt{3}}{2},\frac{3}{2})$ are basis vectors of the triangular Bravais lattice. 

\begin{figure}[t]
\centering
$\begin{array}{c}
   \includegraphics[trim = 5mm 5mm 20mm 20mm, clip, width=6cm]
{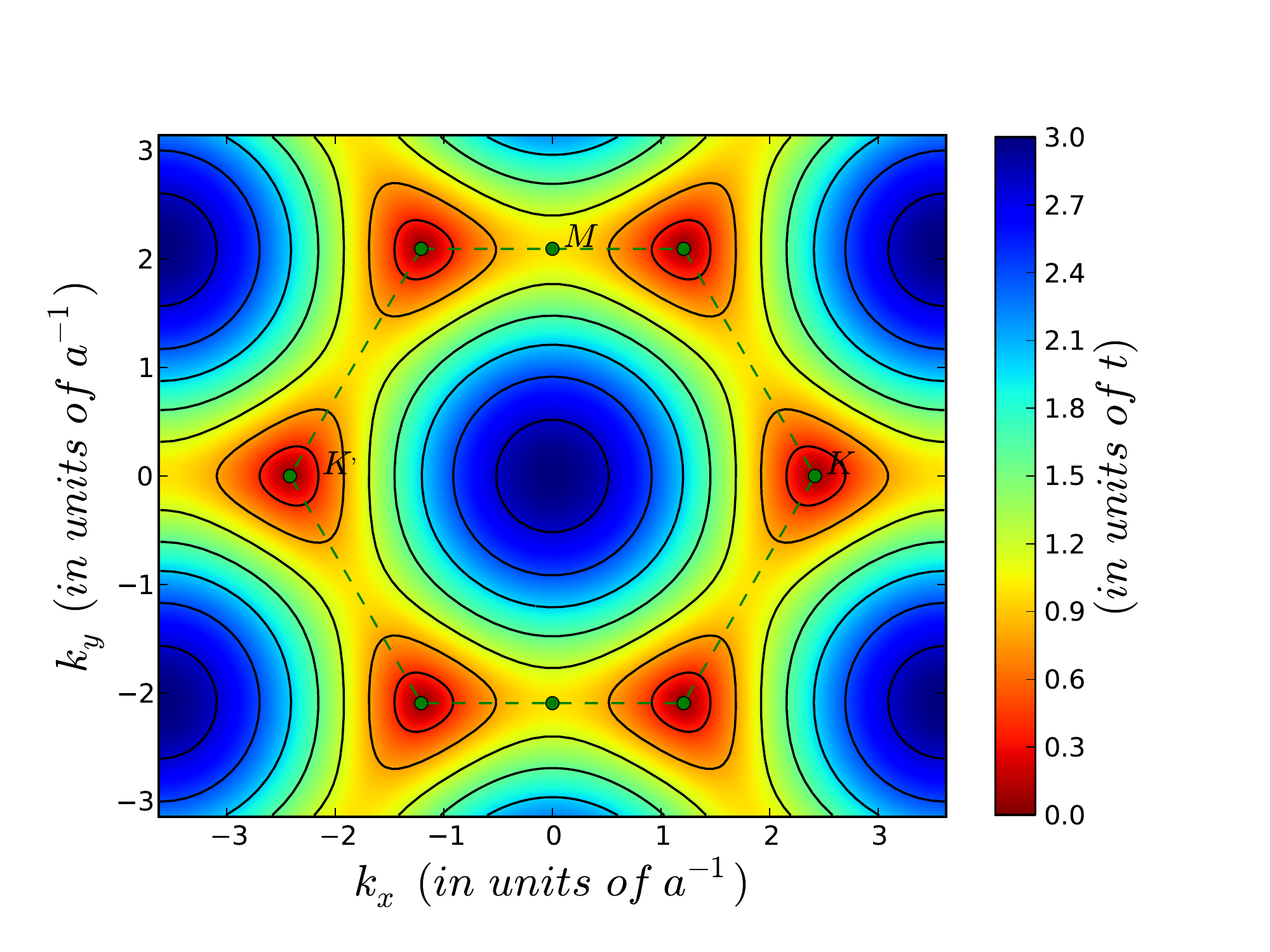}\\
   \includegraphics[trim = 5mm 5mm 20mm 20mm, clip, width=6cm]{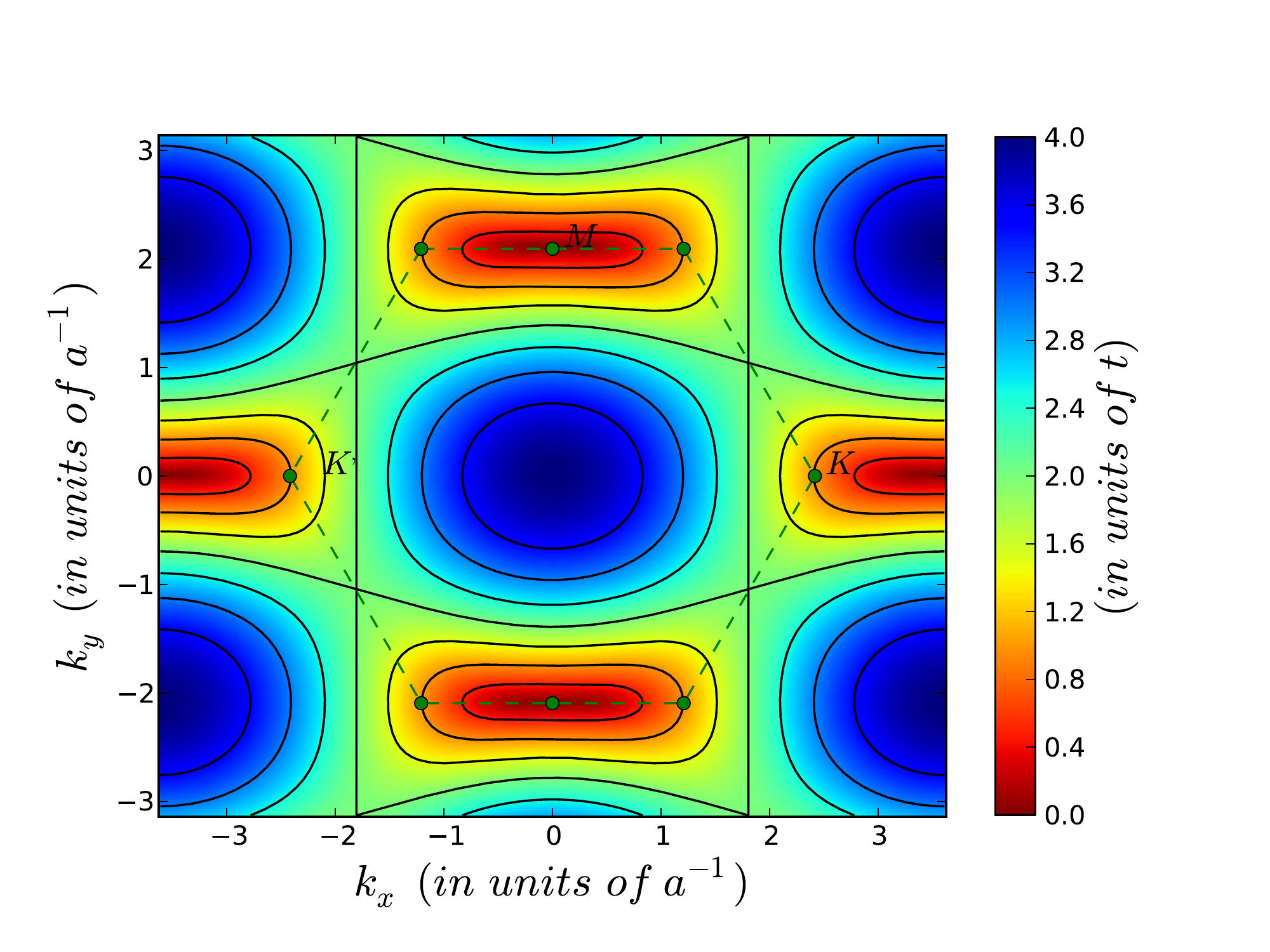}\\
\end{array}$
\caption{Energy spectra when $t'=t$ (top) and $t'=2t$ (bottom). The later corresponds to the merging of Dirac points into a single point  $M$. The green dashed line depicts the Brillouin zone.}
\label{Spectrum}
\end{figure}

The corresponding first Brillouin Zone (BZ) is hexagonal, as depicted by the green dashed line in Fig.~\ref{Spectrum}. Its geometrical properties only depend on the Bravais lattice. Nevertheless the number of atoms per unit cell becomes relevant for the energy spectrum. For graphene (two atoms per unit cell with one electron per atom) the energy bands are well described using a tight-binding model: each $2p_z$ electron may hop between two sites $i$ and $j$ with a given amplitude $t_{ij}$. In this work we only consider the hopping between nearest neighbors, with a fixed hopping amplitude $t\approx2.7$ eV for the nearest neighbor vectors $\vec \delta_1$, $\vec \delta_2$, and a variable amplitude $t'$ for $\vec \delta_3$. The corresponding second-quantized Hamiltonian is given by: 
\begin{align}\label{hamiltonian}
\cal{H}&=\sum_{\langle i,j\rangle}t_{ij}a^\dagger_{i} b_{j}+h.c.\\
&=\int_{BZ} \frac{d^{2}k}{S_{BZ}} \left(a^{\dagger}(\vec k),b^{\dagger}(\vec k) \right) \mathcal{H}_{\vec k} \left(\begin{array}{c}a(\vec k)\\b(\vec k) \\\end{array}\right)
\end{align} 
with $\mathcal{H}_{\vec k}=\left(\begin{array}{cc}0 & f(\vec k)\\f^{*}(\vec k) & 0 \\\end{array}\right)$ \\ 
and $ f(\vec k)=-t\big(e^{-i\vec k.\vec \delta_{1}}+e^{-i\vec k.\vec \delta_{2}}\big)-t'e^{-i\vec k.\vec \delta_{3}}$. $S_{BZ}$ is the area of the BZ. The operators $a$ ($b$) and $a^{\dagger}$ ($b^{\dagger}$) are field operators that respectively annihilate and create an electron on the A (B) sublattice. The energy spectrum is then obtained by diagonalizing the Hamiltonian matrix $\mathcal{H}_{\vec k}$. As there are two atoms per unit cell, there are two energy bands $\epsilon_{\pm}(\vec k)= \pm |f(\vec k)|$. Negative values of $\epsilon$ correspond to the valence band whereas positive ones correspond to the conduction band.  When $t'=t$ there are two inequivalent points $K$ and $K'$ at the corners of the BZ for which the two bands touch. These points are denoted Dirac points since the energy spectrum is conical in their vicinity. Note that the coincidence between the Dirac points (determined by the band structure) and that of the corners of the BZ  (intrinsic to the Bravais lattice) occurs only when $t'=t$. 

Fig. \ref{Spectrum} illustrates the fact that the Dirac points move away from the corners of the BZ when varying the hopping parameter $t'$. Increasing this amplitude from $t$ to $2t$ makes the two inequivalent Dirac points merge \cite{PhysRevB.74.033413, PhysRevB.80.153412} at the $M$ point (right at the middle of edges of the BZ). The critical value $t'=2t$ corresponds to the annihilation of a pair of Dirac points with opposite Berry phases. This topological invariant  changes abruptly from $\pm \pi$ to $0$ at the merging, which thus defines a topological transition between a semi-metallic phase and a band insulator, since a gap opens at the M point for $t'>2t$. This can be seen by expanding $f(\vec k)$ in the vicinity of this point defined by $(0,\frac{2\pi}{3a})$:
\begin{align}\label{fm}
f^{M}(\vec q)&=\left( \Delta+ic_{y}q_{y}+\frac{q_{x}^{2}}{2m^{*}} \right) e^{-i\frac{\pi}{3}} 
\end{align} 
Here $c_y=3ta_{0}$, $2m^{*}=\frac{4}{3ta_{0}^{2}}$, and $\Delta=t'-2t$ characterizes the distance from the topological transition and also gives  the value of the gap when $t'>2 t$. Exactly at the transition ($\Delta=0$), the Hamiltonian exhibits a semi-Dirac energy dispersion such that $\epsilon^{M}_{\pm}(\vec q)= \pm |f^{M}(q)|$ is linear in $q_y$ but quadratic with respect to $q_x$: 
\begin{align}\label{DispersionRelation}
\epsilon^{M}_{\pm}(\vec q)= \pm \sqrt{(c_y q_y)^2 + \left(\frac{q_{x}^{2}}{2m^{*}}\right)^2}
\end{align}

\section{Local density of states in the presence of impurity scattering}
The effects of impurity scattering on the graphene LDOS have been extensively studied\cite{PhysRevB.75.125425} in the past. It has been shown\cite{PhysRevLett.97.226801, PhysRevB.76.165402, PhysRevLett.100.076601} that this gives rise to long-wavelength oscillations that decay as $1/r^2$, instead of the $1/r$ law expected for conventional two-dimensional materials. Here we investigate how the merging of the two Dirac cones changes the form of these long-wavelength oscillations. We start this section with some zero-energy wavefunction arguments along the lines of Ref.~\onlinecite{PhysRevLett.96.036801} that allow us to characterize the impurity state.  Furthermore we perform a more detailed study using analytical (T-matrix approximation) and numerical (tight-binding) techniques.  

\subsection{Wavefunction considerations}
\subsubsection{The sublattice symmetry}
Graphene honeycomb lattice contains two atoms per unit cell ($A$ and $B$), which allows one to define two sublattices. Moreover the Hamiltonian (\ref{hamiltonian}) only takes into account nearest neighbor hopping processes, and neglects hopping between sites belonging to the same sublattice, resulting in a bipartite system. For such systems a generic Hamiltonian takes the form: 
\begin{align}
\mathcal{H}=\left(\begin{array}{cc}0 & T\\ T^{\dag} & 0 \\\end{array}\right)
\label{FiniteHamiltonian}
\end{align}

Without loss of generality, $T$ is a $N_A \times N_B$ block (not necessarily a square matrix), where $N_{A(B)}$ is the number of atoms in the $A(B)$ sublattice, assuming there is only one electron per atom. Here we restrict ourself to $N_B\ge N_A$. Such a Hamiltonian anti-commutes with: 
\begin{align}
\mathcal{S}=\left(\begin{array}{cc} \mathbb I_{N_A} & 0\\ 0 & - \mathbb I_{N_B} \\\end{array}\right)
\end{align}
$\mathbb I_N$ is the $N \times N$ identity matrix so that the unitary operator $\mathcal S$ always squares to $+1$, which defines a chiral symmetry: the sublattice symmetry. 

This fundamental symmetry implies a particle-hole symmetric spectrum, and includes the possibility of existence of zero energy states, which transform into themselves under the transformation $\mathcal{S}$. As a consequence, they have null components on one sublattice. 

Moreover, as pointed out in Ref.~\onlinecite{PhysRevB.34.5208,PhysRevB.49.3190,PhysRevB.66.014204,PhysRevB.77.115109}, every finite bipartite lattice has an extra number of $N_B-N_A$ zero-energy eigenstates living on the sublattice $B$, regardless of the components of the block $T$. This is because the non-zero energy eigenstates appear in pairs: $|\psi\rangle$ and $\mathcal S|\psi\rangle$, and in order to form non-zero energy states, it is necessary to pair a localized state living on the sublattice $A$ with another one living on $B$. As a number of $N_B-N_A$ zero modes living on the sublattice $B$ are unable to satisfy this condition, they are stuck at zero energy, cannot hybridize to $A$ states, and remain localized purely on $B$. 

\subsubsection{Zero-energy impurity wavefunction}
\begin{figure}[t]
\centering
\includegraphics[width=7cm]{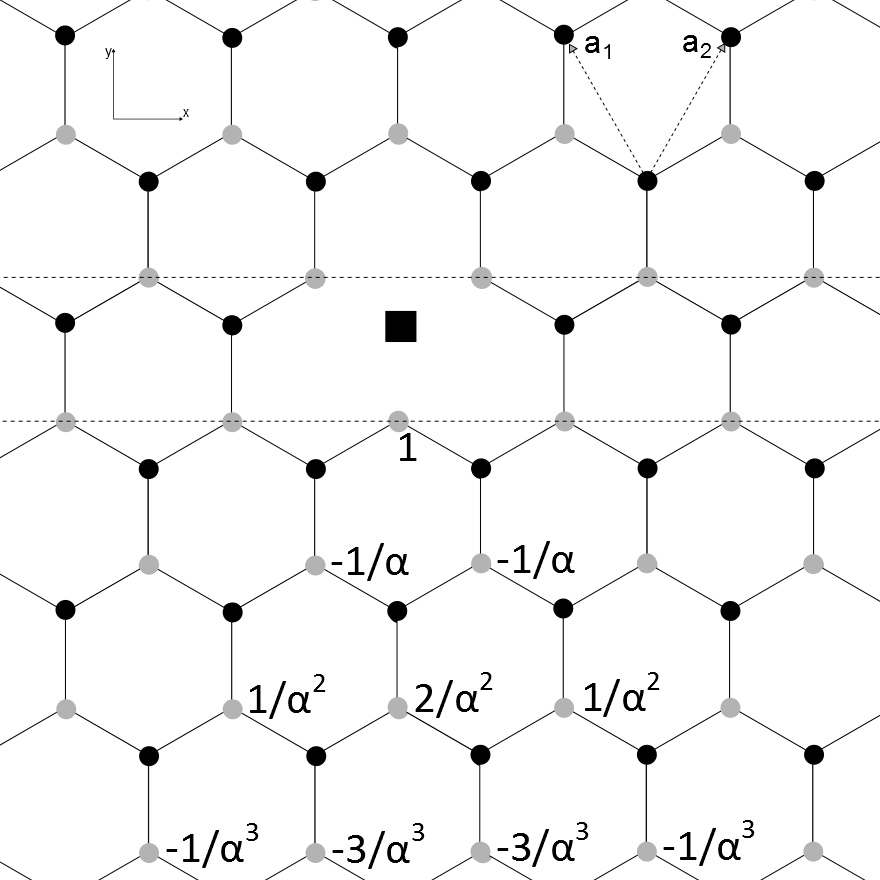}
\caption{The zero-energy wavefunction components for $t'=\alpha t$ with $\alpha \ge2$. The wavefunction is zero for all sites for which no value is specified. The black square denotes the vacancy. The direction of anisotropy is the $y$-direction. The two dashed lines are the boundaries of the upper and respectively lower half-planes.}
\label{Handmade}
\end{figure}

In the presence of a single vacancy, $N_B-N_A=1$, and we have a single zero-mode impurity-state wavefunction. Here the fundamental point is that varying the parameter $t'$ does not change the structure of the matrix (\ref{FiniteHamiltonian}). Then the sublattice symmetry ensures that such a zero-mode does exist, even in the gapped phase ($t'>2$). As a consequence, this zero-energy state is a good candidate to characterize the Dirac cones merging in real space. In this section we study the form of its wavefunction, using simple arguments along the lines of Ref.~\onlinecite{PhysRevLett.96.036801}. We already know that such a wavefunction has null components on the $A$ sites, represented by the black disks in Fig.\ref{Handmade}, and we need to determine its value on the $B$ sublattice. In Ref.~\onlinecite{PhysRevLett.96.036801}, the authors have determined the exact analytic form of the impurity wavefunction for an isotropic honeycomb lattice with a single vacancy. Their method consists in an appropriate matching of the zero modes of two semi-infinite and complementary graphene sheets. This is the method we generalize in what follows for anisotropic graphene.

In Fig.~\ref{Handmade} the two semi-infinite graphene sheets are defined such that their edges are orthogonal to the anisotropic direction, along which $t'=\alpha t$ with $\alpha \ge 2$. Here we have introduced an anisotropy parameter $\alpha$, ($\alpha=2$ exactly at the merging) that allows us to explore the gapped phase beyond the merging point. The upper half-plane has a `bearded' edge (as indicated by the upper dashed line in Fig.~\ref{Handmade}), whereas the lower half-plane has a zigzag edge. 

Let us first consider the lower half-plane terminated by the zigzag edge. The form of the edge states for a semi-infinite zigzag ribbon is well known \cite{JPSJ.65.1920, PhysRevB.54.17954} for isotropic graphene. In a manner similar to that of Ref.~\onlinecite{JPSJ.65.1920}, the edge states for anisotropic graphene can be determined by imposing the condition: $|2\cos(k/2)|\le \alpha$, where $k$ is the momentum along the edge. While for isotropic graphene ($\alpha=1$) this condition is verified for $2\pi/3\le k\le 4\pi/3$, above the merging point ($\alpha \ge 2$) such a condition is satisfied for all values of $k$, $0\le k\le 2\pi$. Next, regarding the complementary semi-infinite bearded plane, the condition becomes: $|2\cos(k/2)|\ge \alpha$, which cannot be satisfied for any $k$ when $\alpha > 2$. The case $\alpha=2$ leads to $k=0$, associated to an extended state, and there are no allowed edge states in this limit. 

The condition that the impurity wavefunctions on the two semi-infinite planes match at the interface can be written as: 
\begin{align}
\alpha b_{m,0}^{(l)} + b_{m,0}^{(u)} + b_{m+1,0}^{(u)}=0
\label{EdgeRelation}
\end{align}
where $b_{m,n}^{(l)}$ ($b_{m,n}^{(u)}$) corresponds to a given site of the lower (upper) half plane characterized by $\vec{r}_{m,n}=m(\vec a_2-\vec a_1)-n\vec a_1$. The origin is defined to be on the $B$ atom right below the vacancy in Fig. \ref{Handmade}. The above relation is valid everywhere on the edges except for $m=0$. Introducing $b_{m,0}=\sum_k b_{k,0} e^{ikm}$, the condition (\ref{EdgeRelation}) can be rewritten in terms of momentum as:
\begin{align}
\alpha \sum_k b_{k,0}^{(l)} e^{ikm}+ \sum_{k'} b_{k',0}^{(u)} (1+e^{ik'}) e^{ik'm}=0.
\label{boundary}
\end{align}
A possible solution for the boundary conditions is $b_{k,0}^{(l)}=1$ with $0\le k\le 2\pi$  and $b_{k',0}^{(u)}=0$. As for the case of isotropic graphene studied in Ref.~\onlinecite{PhysRevLett.96.036801}, this corresponds to the edge solutions for two isolated complementary semi-infinite planes. Considering the lattice as infinite, the discrete sum in (\ref{boundary}) turns into an integral, and the impurity wavefunction can be written as:
\begin{align}
\label{ZeroModeWF}
b_{m,n}^{(l)}&\sim \int_0^{2\pi}dk (-2/\alpha)^{n} \cos^{n}(k/2) e^{ik(m+\frac{n}{2})} \notag\\
	&\sim (-1)^{n}\frac{e^{-n ln(\frac{\alpha}{2})-\frac{(2m+n)^2}{2n}}}{\sqrt n}
\end{align}
This approximation is valid for large distances when defining $x=a_0\sqrt{3}(2m+n)/2$ and $y=-n3a_0/2$. Most useful to compare with the results of the subsequent sections is the behavior of the wavefunction along the direction $x=0$. Along this direction, the zero energy impurity state exhibits an exponential decay with the distance from the impurity in the gapped phase ($\alpha>2$), whereas it decays as $1/\sqrt{y}$ at the merging ($\alpha=2$). Moreover, if one evaluates the impurity wavefunction in the semi-metallic phase for $1<\alpha<2$, one can check that it still decreases as $1/r$ in both directions, albeit exhibiting a strong asymmetry between $x$ and $y$. Hence the decay law of the zero-energy impurity states provides a real-space signature of the Dirac cone merging.

Furthermore we can evaluate the amplitude of this impurity state by hand, see Fig.~\ref{Handmade}, by searching for a decaying wavefunction with null components everywhere in the semi-infinite bearded ribbon. The condition (\ref{EdgeRelation}) becomes $\alpha b_{m,0}^{(l)}+0+0=0$ and must be satisfied at each $A$ site between the two dashed lines, except for the impurity site. So the wavefunction has zero components along the zig-zag edge, except at the site situated right under the impurity, for which we take $b_{0,0}^{(l)}=1$. Then the Hamiltonian (\ref{FiniteHamiltonian}) implies that $\alpha b_{m,1}^{(l)}+ b_{m,0}^{(l)}+ b_{m+1,0}^{(l)}=0$ for all values of $m$. This leads to $b_{-1,1}^{(l)}=b_{0,1}^{(l)}=-1/\alpha$ and $b_{m,1}^{(l)}=0$ for all other sites with $n=1$. If we extend this analysis to the subsequent rows, we obtain the impurity wavefunction values shown in Fig.~\ref{Handmade}. So above the merging point, this peculiar localized state describes electrons that are localized only in the lower-half plane of the graphene sheet with a single impurity.

\begin{figure}[t]
\includegraphics[trim = 5mm 10mm 10mm 10mm, clip, width=7cm]
{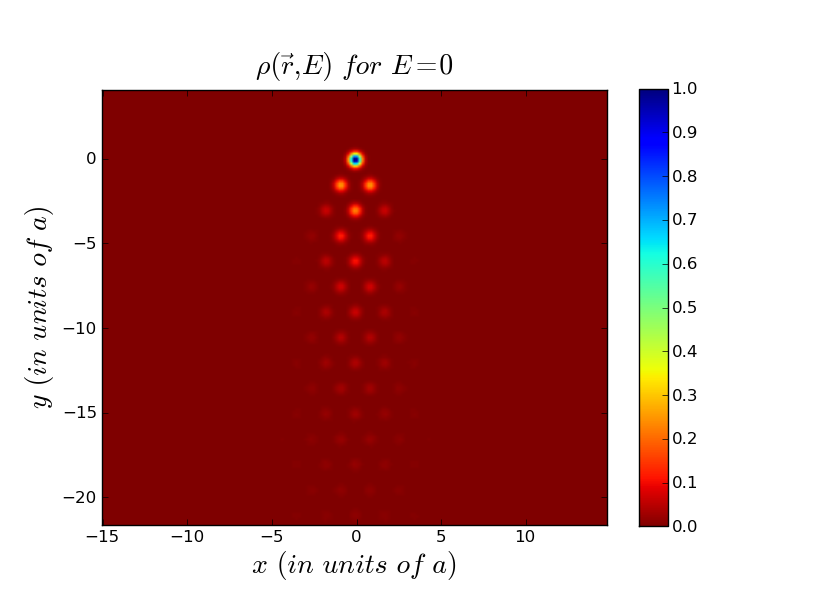}
\caption{\small Snapshot of the LDOS obtained using exact diagonalization. We plot the zero-energy impurity state slightly above the merging point, ($\alpha=2.1$) when a small gap opens in the spectrum. The highest-intensity site (in blue) corresponds to the $B$ site right below the impurity.}
\label{GappedSpectrum}
\end{figure}

\subsection{Exact diagonalization}
In order to obtain the local density of states on the lattice in the presence of disorder one can diagonalize exactly or using numerical approximation the lattice tight-binding Hamiltonian. Here, since the systems we consider are not too large (around 1800 atoms) we base ourselves on an exact diagonalization technique. The lattice Hamiltonian is defined by: 
\begin{align}\label{TightBindingHamiltonian}
\cal{H}&=-\sum_{\langle i,j\rangle}t_{ij}|i \rangle \langle j|+V_0 |0 \rangle \langle 0|
\end{align} 
where $|i \rangle$ stands for the $2p_z$ non-hybridized orbital centered on site $i$. The impurity that we consider is a vacancy, which can be modeled by removing the corresponding atom from the lattice, or by taking an infinite value for $V_0$. We denote by  $|k\rangle$ the eigenstate corresponding to the eigenvalue $E_k$. The eigenfunctions of (\ref{TightBindingHamiltonian}) can be written as a linear combination of individual orbitals:  

\begin{eqnarray}
\left| k \right\rangle =\sum\limits_{i}{{{c}_{ki}}\left| i \right\rangle},\\
{{c}_{ki}}=\left\langle  i | k \right\rangle
\end{eqnarray}

The LDOS, corresponding to the number of available states on a site $i$ at energy $E$  is then given by: 
\begin{align}\label{TB-Ldos}
\rho_i(E)&=
\sum_k {{\left| {{c}_{ki}} \right|}^{2}}  f_k(E)
\end{align} 
where $f_k(E)=\delta(E-E_k)$ is the Dirac delta function centered on the eigenenergy $E_k$. While in an infinite system this procedure automatically yields a continuous energy spectrum, in a finite sample the spectrum is smoothed by taking $f_k$ to be a Gaussian or a Lorentzian. 

In Fig.~\ref{GappedSpectrum} we show the LDOS obtained using this method at zero energy in the gapped phase ($\alpha=2.1$). This is in agreement with the zero-energy wavefunction described previously and depicted in Fig.~\ref{Handmade}. The result for the spatial dependence of the LDOS at a finite energy is presented in Fig.~\ref{Analytic}. Note the strong asymmetry of the LDOS between the positive and negative values of $y$ close to the impurity. While some of these features are consistent with the previous observations concerning the impurity-state wavefunctions, we also investigate them in more details (for example in what concerns their energy dependence) in the next section, via the T-matrix approximation technique. 


\subsection{T-matrix approximation}
\begin{figure}[t]
\centering
\includegraphics[width=7cm]{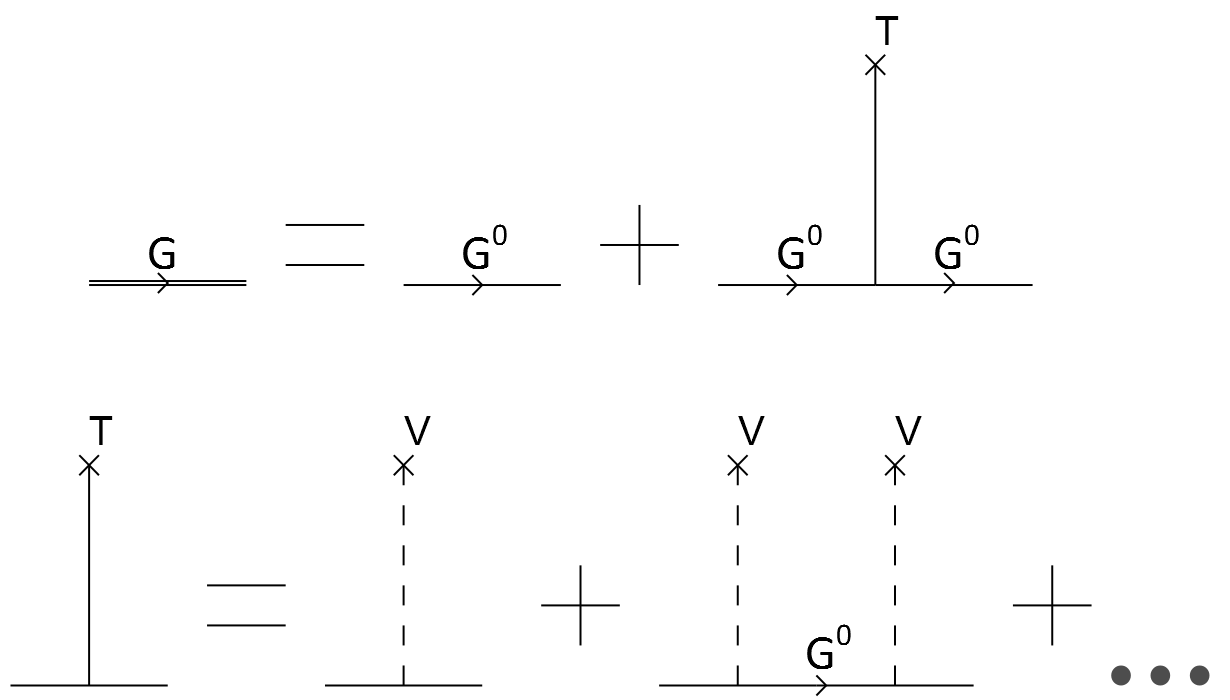}
\caption{Diagrammatic perturbative expansion of the generalized Green's function to all orders in the impurity potential.}
\label{Diagrams}
\end{figure}

The T-matrix approximation consists in a perturbative expansion of the Green's function to all orders in the impurity scattering, as shown in Fig.~\ref{Diagrams}. In this paper we consider the case of a localized impurity $V(\vec r)=V_0 \delta(\vec r)$ situated on a sublattice $A$, for which  $V(q)$ is independent of momentum. We consider as impurity a vacancy, for which $V_0$ becomes infinite. 

The expansion of the T-matrix in Fig.~\ref{Diagrams} is a geometric series and the infinite summation of diagrams can be performed exactly:
\begin{align}\label{TmatrixExpression}
T(i\omega_n)=[I_2-V\int_{BZ} \frac{d^{2}k}{S_{BZ}}G_0(\vec k,i\omega_n)]^{-1}V
\end{align} 
where $S_{BZ}$ is the area of the BZ, $i\omega_n$ a Matsubara frequency,  $G_0(\vec k,i\omega_n)=[i\omega_n I_2-\mathcal{H}_{\vec k}]^{-1}$ is the unperturbed Green's function, $I_2$ the $2\times 2$ identity matrix, and $V=\left(\begin{array}{cc}V_0 & 0\\0 & 0 \\\end{array}\right)$.

We define $\Delta\rho$ as the correction to the LDOS due to the impurity. According to Fig.~\ref{Diagrams}, we have: 
\begin{align}
\Delta G(\vec Ri, \vec Rj, E)&\doteq G(\vec Ri, \vec Rj, E)-G^0(\vec Ri-\vec Rj, E) \notag\\
	&=G^0(\vec Ri,E)T(E)G^0(-\vec Rj, E)
\end{align} 
The correspondence between the components of $\Delta G$  in the continuum and on the lattice is the following: 
\begin{align}\label{GFcontinuumLattice}
\Delta{G}_{\alpha \beta}(\vec r_{1}, \vec r_{2},E) &= \notag \\ 
 \sum_{i,j} \phi_{\beta}(\vec r_{1} - \vec R_{j})&\phi^{*}_{\alpha}(\vec r_{2} - \vec R_{i}) \Delta{G}_{\alpha \beta}(R_{j}, R_{i},E),
\end{align} 
where $\phi_{\alpha(/\beta)}$ is a carbon $2p_z$ orbital, the $\alpha$ and $\beta$ indices denote the sublattice, whereas $i$ and $j$ label the unit cell.
The impurity correction to the LDOS is given by: 
\begin{align}\label{LDOSdefinition}
\Delta\rho(\vec r,E)=-\frac{1}{\pi}Im\big[Tr\{\Delta G(\vec r, \vec r, E)\}\big]
\end{align} 
which yields in the momentum space:
\begin{align}\label{kDOSdefinition}
\Delta\rho(\vec q,E) & =  \\ 
\frac{i}{2\pi} \int_{BZ}&\frac{d^{2}k}{S_{BZ}} Tr\{\Delta G(\vec k + \vec q,\vec k,E)-\Delta G^{*}(\vec k,\vec k+\vec q,E)\} \notag
\end{align} 

\subsubsection{Momentum dependence of the Fourier transform of the LDOS}
\begin{figure*}[h]
$\begin{array}{ccc}
   \includegraphics[trim = 25mm 0mm 25mm 0mm, clip, width=5.5cm]{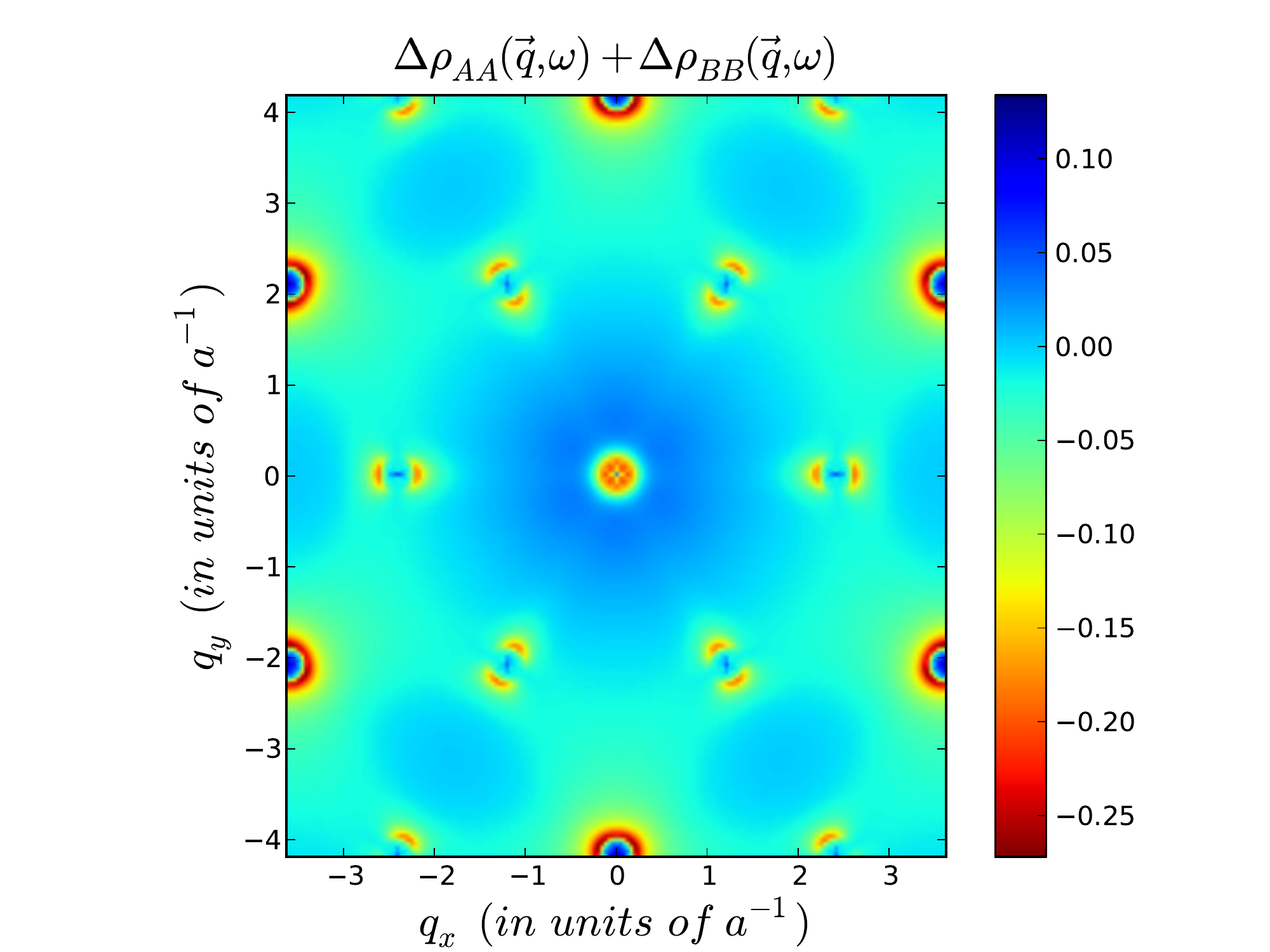}&  
   \includegraphics[trim = 25mm 0mm 25mm 0mm, clip, width=5.5cm]{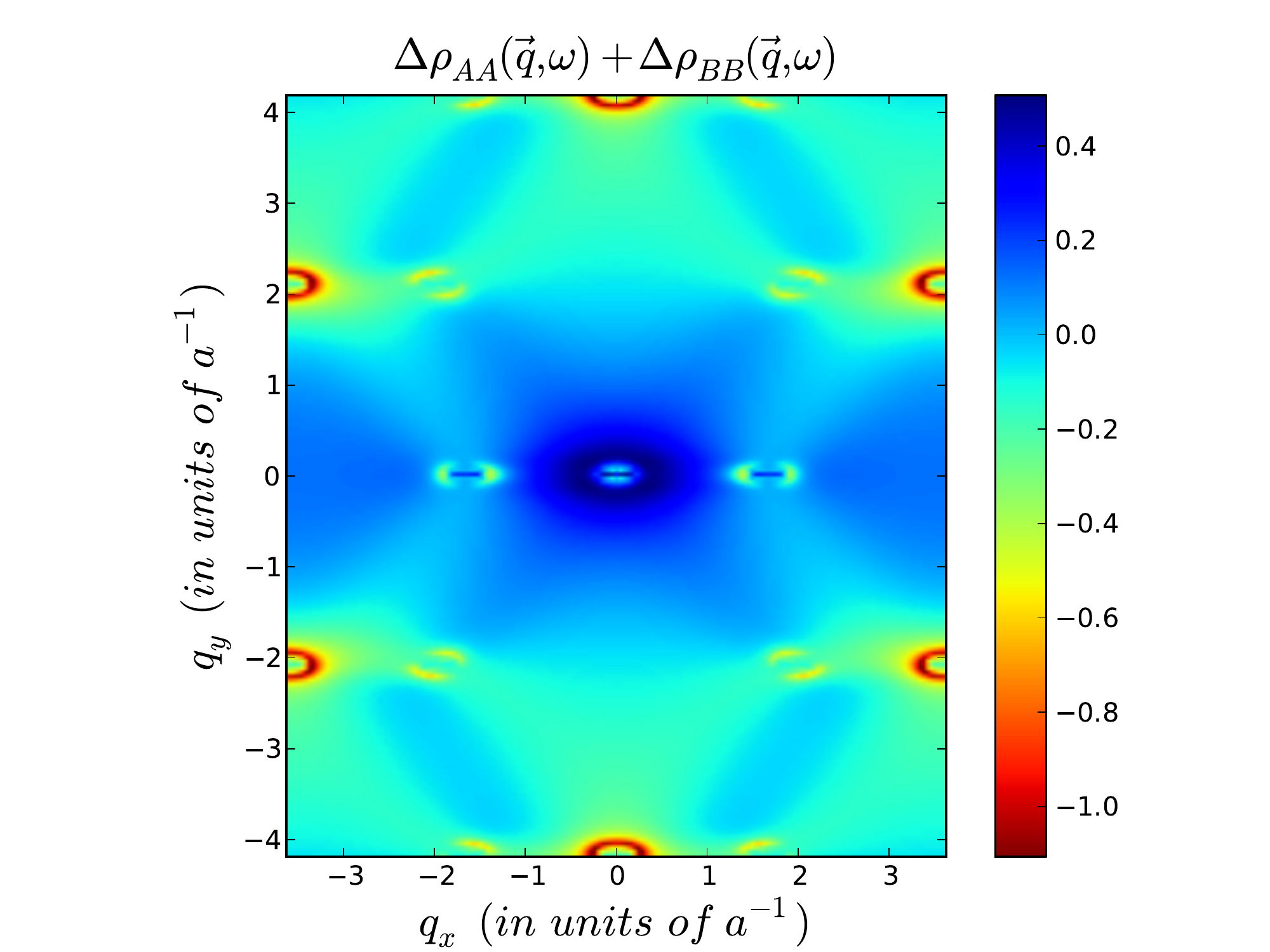}&
   \includegraphics[trim = 25mm 0mm 25mm 0mm, clip, width=5.5cm]{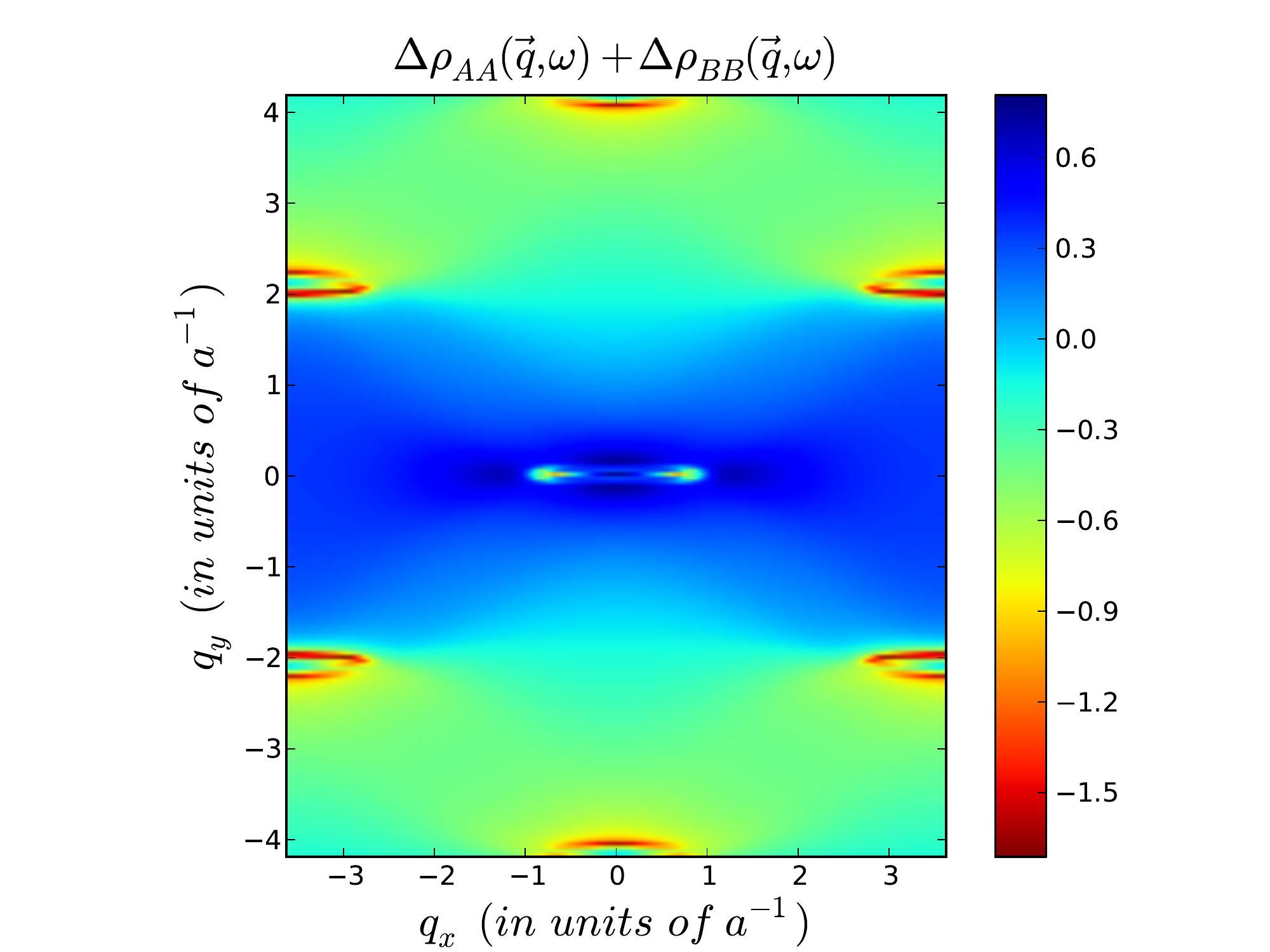}\\
   \includegraphics[trim = 25mm 0mm 25mm 0mm, clip, width=5.5cm]{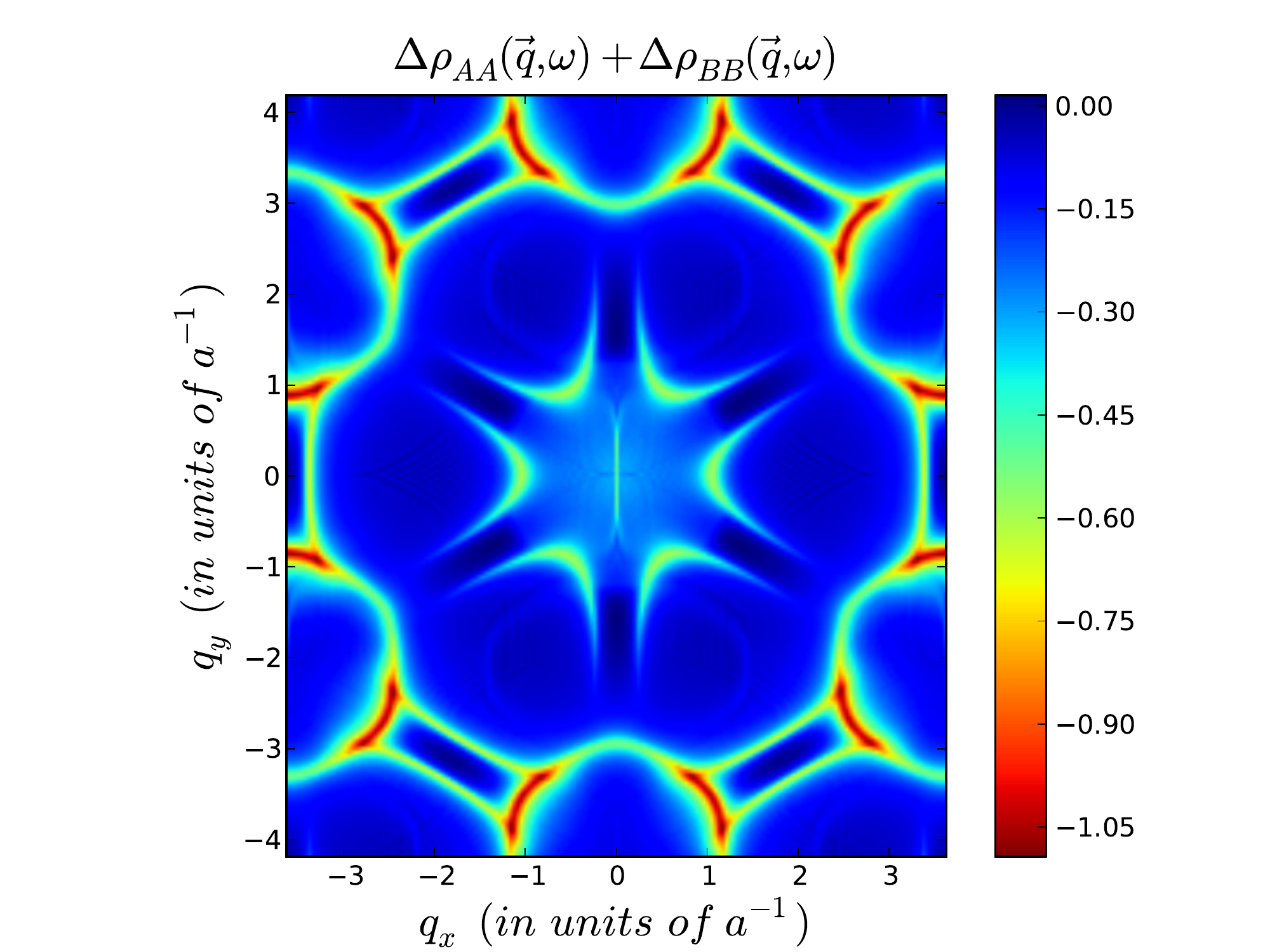}&
   \includegraphics[trim = 25mm 0mm 25mm 0mm, clip, width=5.5cm]{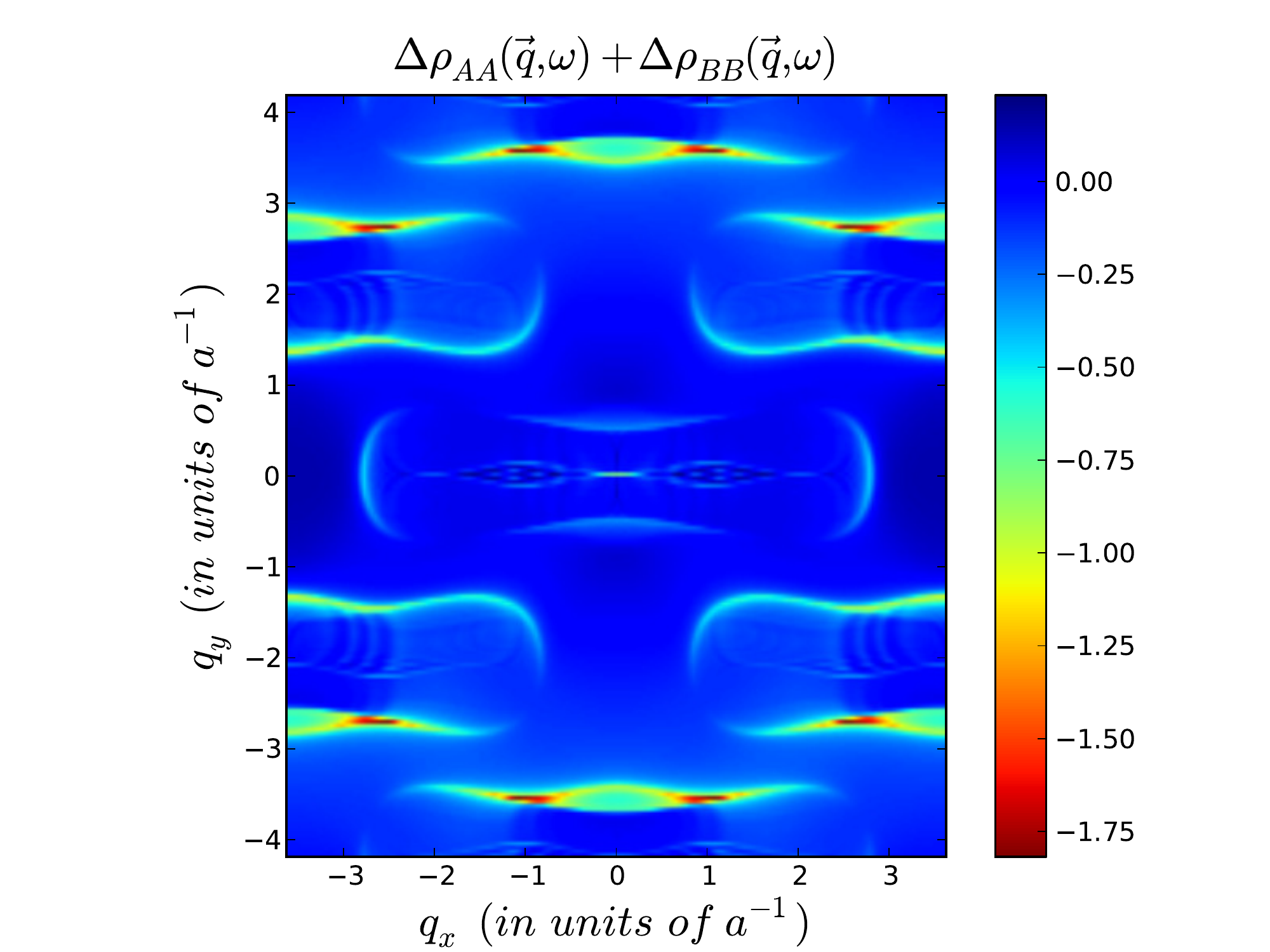}&
   \includegraphics[trim = 25mm 0mm 25mm 0mm, clip, width=5.5cm]{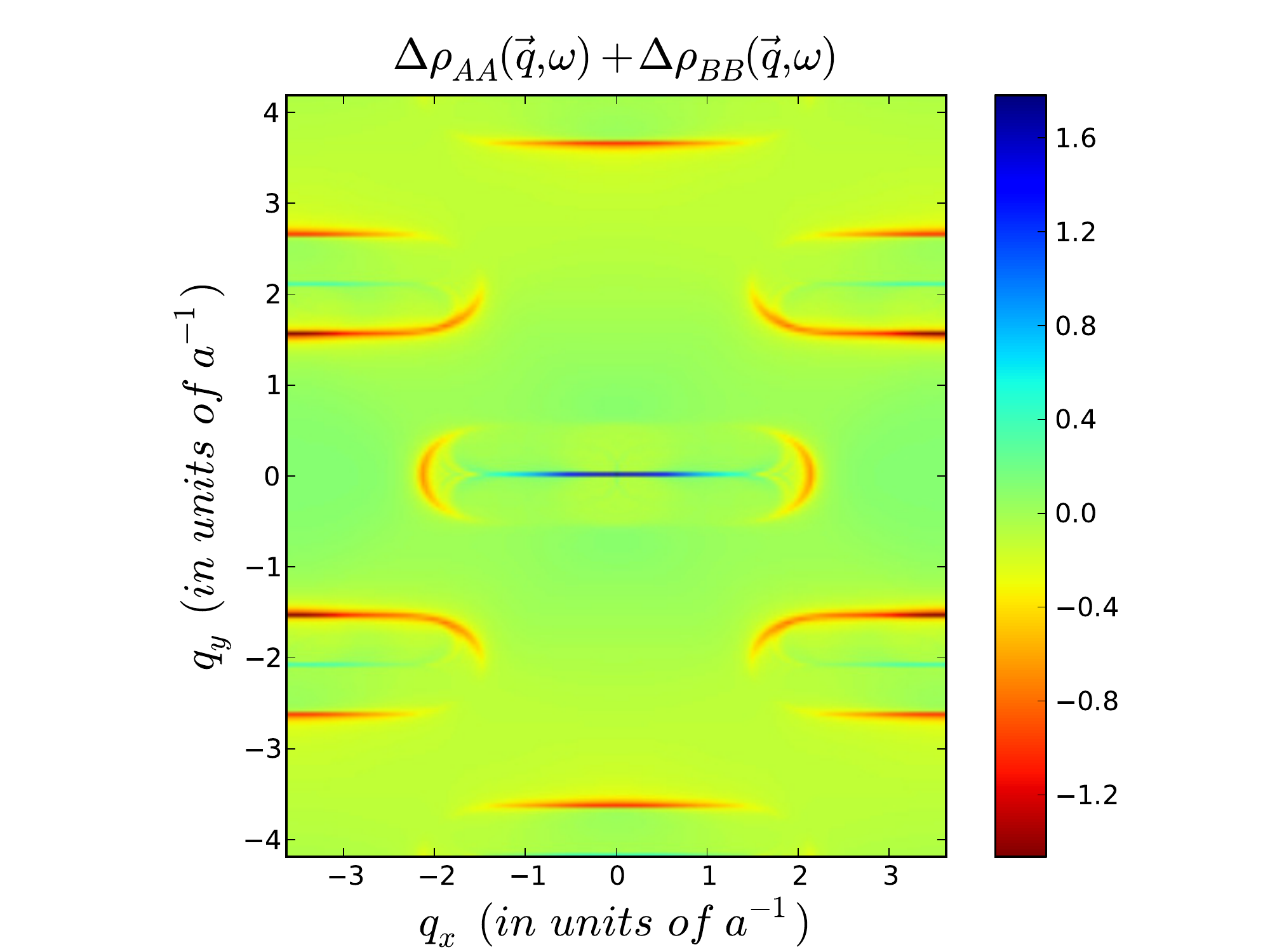}\\
\end{array}$
\caption{\small Fourier transform of the LDOS correction $\Delta \rho$ when $t'=t$ (first column), $t'=1.5 t$ (second column) and $t'=2t$ (third column). The energy is $\omega=0.15t$ for the first row and $\omega=0.80t$ for the second row.}
\label{kDOS}
\end{figure*}

We focus first on the evaluation of the momentum dependence of $\Delta \rho$, corresponding to the measured Fourier transform of the LDOS. In Fig.~\ref{kDOS}  we plot this momentum dependence for $\omega=0.15 t$ and $\omega=0.8t$. The first column corresponds to isotropic graphene ($t'=t$). As noted previously  \cite{PhysRevLett.100.076601} the central circle (in red) corresponds to intranodal scattering, whereas the outer regions around the corners of the BZ correspond to internodal scattering. 
In the second column we consider an intermediate value of $t'=1.5 t$, while in the third column we consider the Dirac-cone merging limit $t'=2t$. We note that the outer regions disappear at the merging point for which internodal quasiparticle scattering no longer exists. Moreover, we note that the features corresponding to intranodal scattering, centered on the sites of the reciprocal lattice are strongly anisotropic, corresponding to the low-energy anisotropic semi-Dirac spectrum.
\begin{figure*}[h]
$\begin{array}{ccc}
   \includegraphics[trim = 5mm 5mm 20mm 10mm, clip, width=6cm]{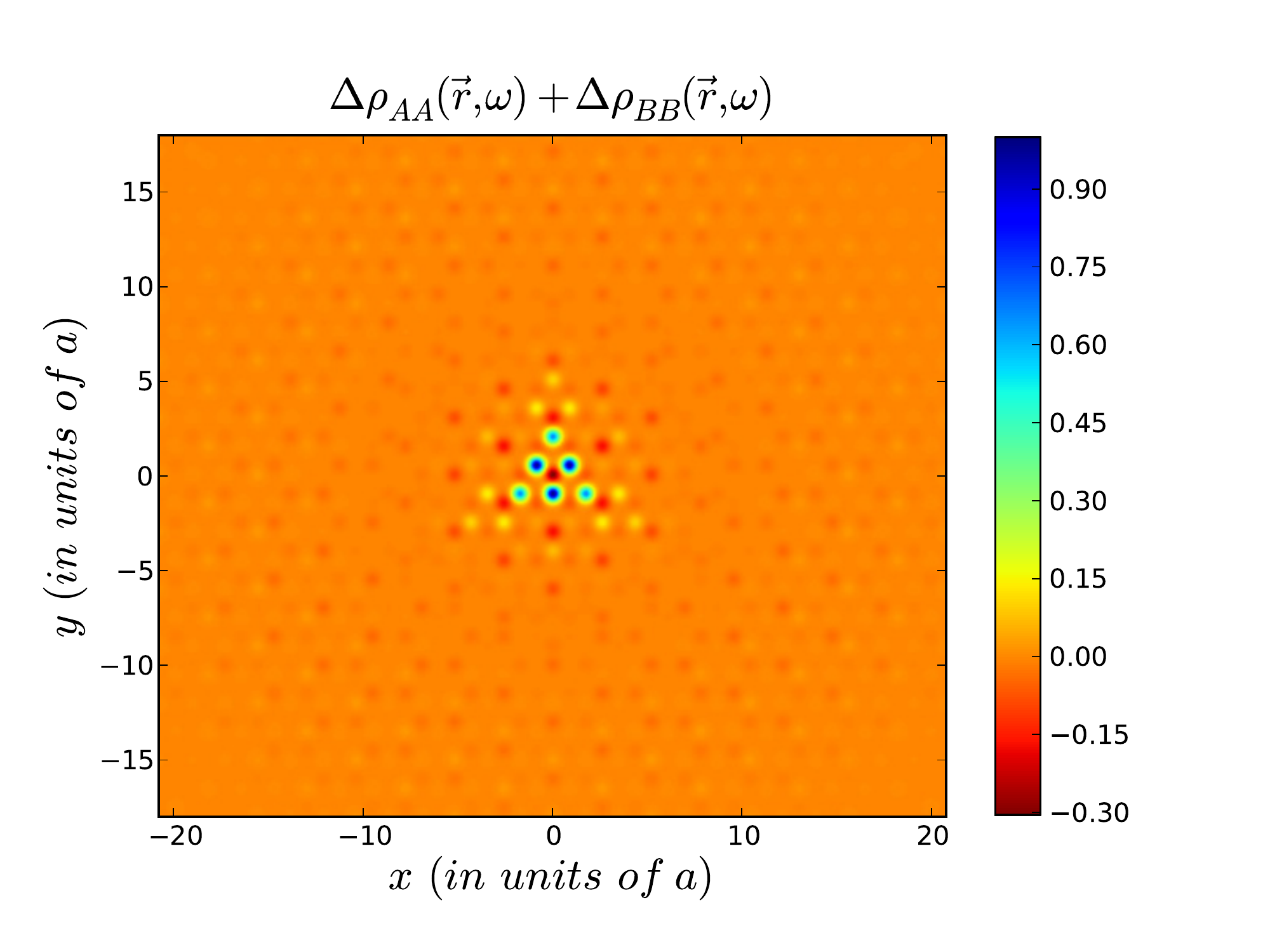}&
   \includegraphics[trim = 5mm 5mm 20mm 10mm, clip, width=6cm]{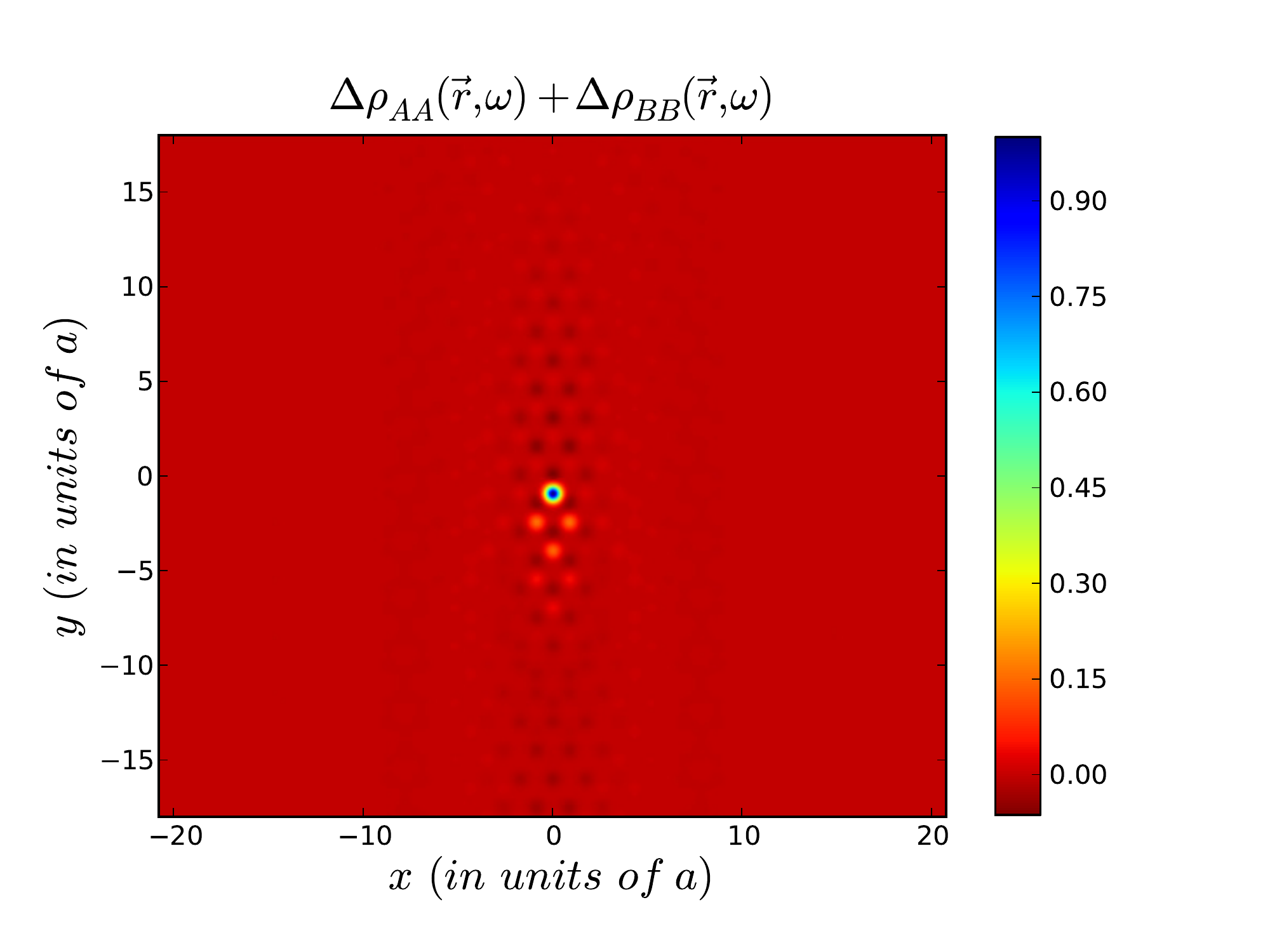}&
   \includegraphics[trim = 5mm 5mm 20mm 10mm, clip, width=6cm]
{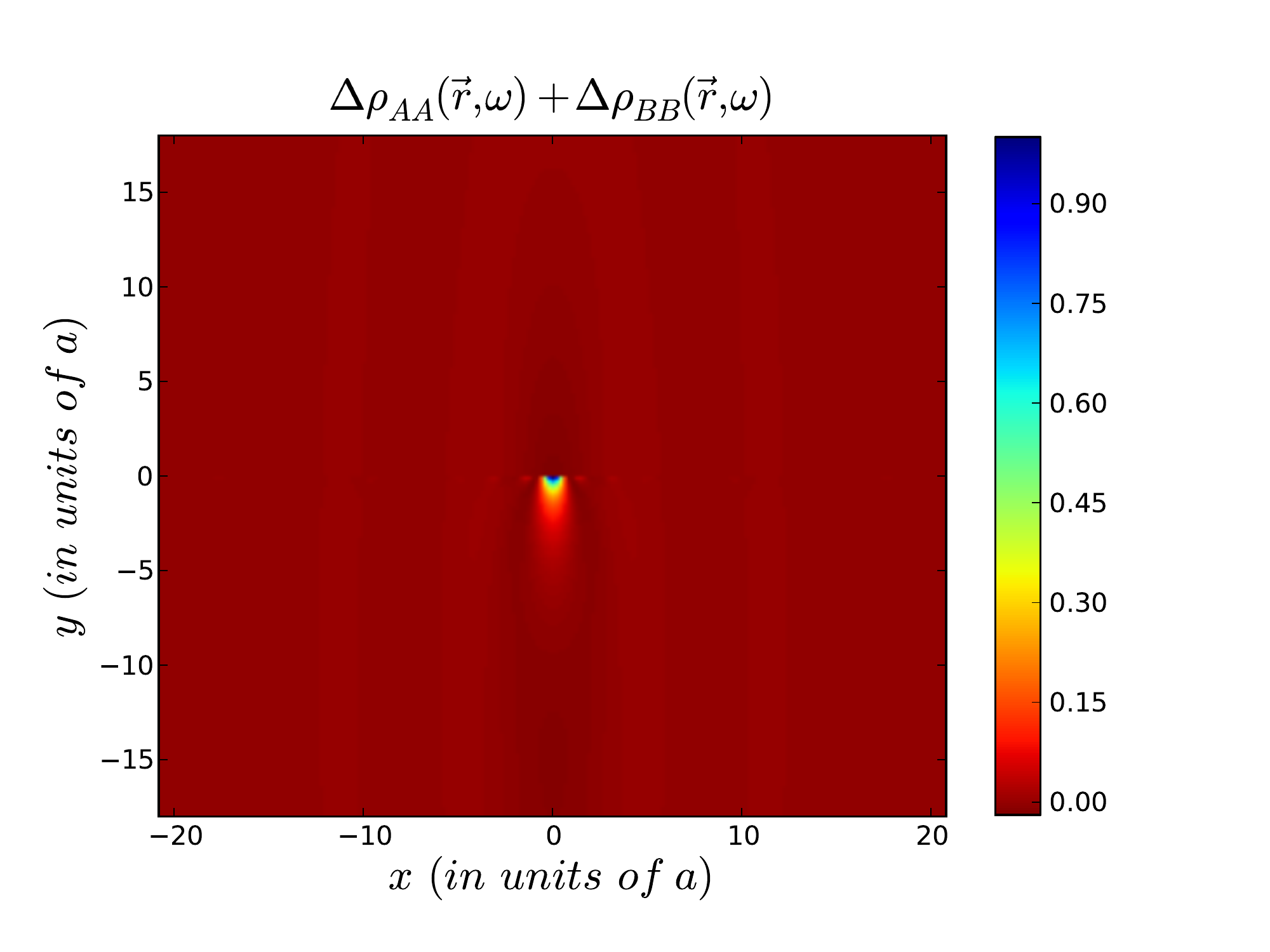}\\
\end{array}$
\caption{\small Correction of the LDOS $\Delta \rho$ (in arbitrary units) for $t'=t$ (first column), $t'=2t$ (second and third columns), obtained using the full T-matrix approximation (first and second columns) and the low-energy expansion (third column). The energy is $\omega=0.15t$.}
\label{LDOS}
\end{figure*}

\subsubsection{Friedel oscillations in real space}
In what follows we focus on the real space form of the Friedel oscillations. While they can be evaluated numerically for arbitrary energy and position using the formulae presented in the previous section, we can also obtain an analytical form of these oscillations in certain limits by performing an expansion of the Hamiltonian at low energy, where the physics is dominated by the semi-Dirac spectrum around the $M$ point. Using the expansion (\ref{fm}) with $\Delta=0$, the unperturbed Green's function for $x=0$ can be rewritten as: 

\begin{align}\label{GreenFunctionMatrix}
G^{0}(0,y,\omega)=\left(\begin{array}{cc}G^{0}_{AA}(0,y,\omega) & G^{0}_{AB}(0,y,\omega) \\ G^{0}_{BA}(0,y,\omega) & G^{0}_{BB}(0,y,\omega) \\\end{array}\right)
\end{align} 
with, 
\begin{align}\label{GreenFunctionExpression}
G^{0}_{AA}(0,y,\omega)&=-i2^{-5/4}\pi^{3/2}\Gamma(1/4)\omega^{-1/4}y^{1/4}H^{(1)}_{-1/4}(\omega y) \notag \\ 
G^{0}_{BB}(0,y,\omega)&=G_{AA}(0,y,\omega) \notag \\
G^{0}_{AB}(0,y,\omega)&=-iA2^{-3/4}\pi^{3/2}\Gamma(3/4)\omega^{1/4}y^{-1/4}H^{(1)}_{1/4}(\omega y) \notag \\ 
			&\mp iAi2^{-5/4}\pi^{3/2}\Gamma(1/4).\omega^{3/4}y^{1/4}H^{(1)}_{3/4}(\omega y) \notag \\
G^{0}_{BA}(0,y,\omega)&=-i\bar{A}2^{-3/4}\pi^{3/2}\Gamma(3/4)\omega^{1/4}y^{-1/4}H^{(1)}_{1/4}(\omega y) \notag \\ 
			&\mp i\bar{A}i2^{-5/4}\pi^{3/2}\Gamma(1/4)\omega^{3/4}y^{1/4}H^{(1)}_{3/4}(\omega y) 
\end{align} 
where $H^{(1)}_{\nu}$ are Hankel functions of the first kind, $\Gamma$ is the Euler gamma function and $\bar{A}$ the conjugate of an arbitrary phase factor $A$ that depends on the basis we choose to write $f(\vec k)$; the value of any observable physical quantity should be independent of this choice\cite{2009NJPh...11i5003B}. Note that on the right-hand-side of the above formulae we have chosen to denote the absolute value $|y|$ simply by $y$. Moreover, the $\mp$ signs correspond to a positive and respectively negative value for $y$. The antisymmetric form of $G_{AB}$ and $G_{BA}$ is responsible for an anisotropy of the impurity-induced corrections to the LDOS on the $B$ sublattice, as it will be described in more detail in what follows.  According to Eq.~(\ref{LDOSdefinition}), the LDOS correction on each sublattice is given by: 
\begin{align}\label{LatticesLDOSFormula}
&\Delta \rho_{AA}(0,y,\omega)=-\frac{1}{\pi}Im\big[G^{0}_{AA}(0,y,\omega)t(\omega)G^{0}_{AA}(0,-y,\omega)\big] \notag\\
&\Delta \rho_{BB}(0,y,\omega)=-\frac{1}{\pi}Im\big[G^{0}_{BA}(0,y,\omega)t(\omega)G^{0}_{AB}(0,-y,\omega)\big]
\end{align} 
where $t(\omega)$ is the only non-nul component of the T-matrix. In the case of an infinite impurity potential (vacancy), it takes the form:
\begin{align}
\label{Tcomponent}
t(\omega)\sim \frac{e^{-i\pi/4}}{\omega^{1/2}}
\end{align}

At this point, we can check that at zero energy our T-matrix calculations recover the same expression for the LDOS as the one obtained from the zero-energy wavefunction considerations. In the limit $\omega \to 0$: 
\begin{align}
\label{ZeroEnergyGreen}
&G^{0}_{AA}(0,y,\omega)\sim \omega^{1/2} \notag\\
&G^{0}_{AB}(0,y,\omega)\sim A \frac{e^{i\pi}}{\sqrt{y}}\theta(y)
\end{align} 
where $\theta$ is the Heaviside step function.
The LDOS correction then vanishes on the sublattice A. We stress that the sublattice symmetry implies that at zero energy the LDOS on the sublattice A is zero, whereas it behaves in the following manner on the sublattice B:
\begin{align}\label{ZeroEnergyLDOS}
\Delta \rho_{BB}(0,y,\omega)\sim \frac{\theta(-y)}{\omega^{1/2}y}
\end{align} 
This result is in agreement with the analysis of the zero-energy wavefunction. Remember that the impurity wavefunction decays as $1/\sqrt{y}$ with the distance from the impurity  (cf. (\ref{ZeroModeWF})), which then leads to a $1/y$ decay for the LDOS. 

\begin{figure*}[h]
\centering
$\begin{array}{ccc}
\includegraphics[trim = 15mm 0mm 15mm 0mm, clip, width=5.5cm]{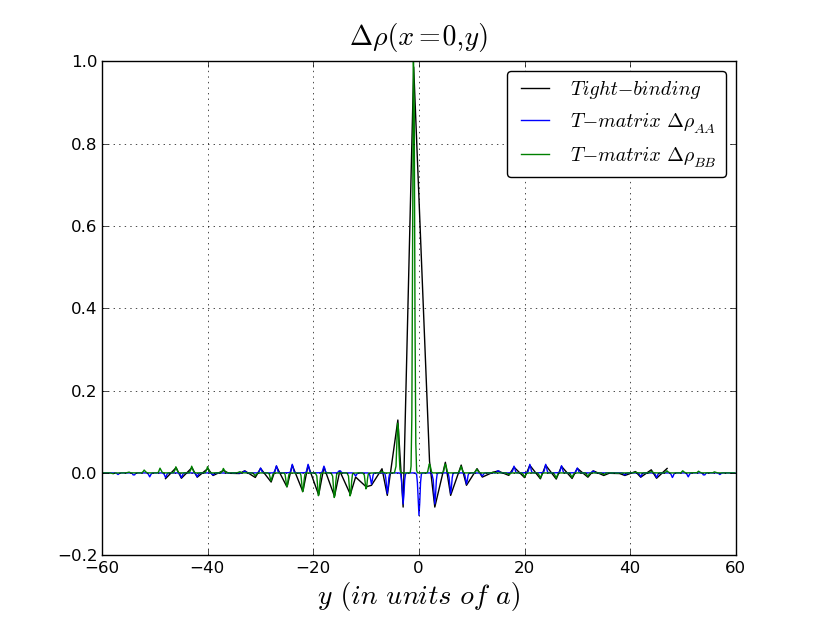}&
\includegraphics[trim = 15mm 0mm 15mm 0mm, clip, width=5.5cm]{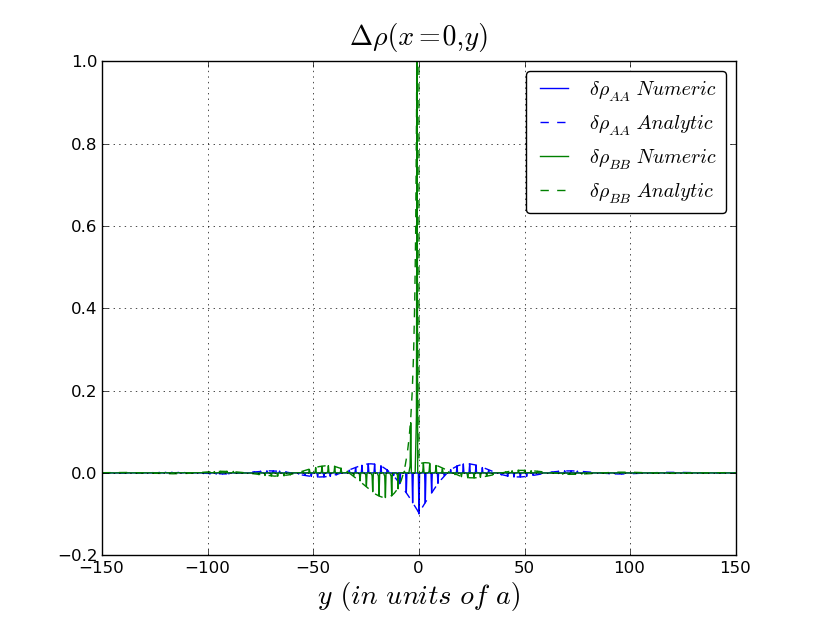}&
\includegraphics[trim = 15mm 0mm 15mm 0mm, clip, width=5.5cm]{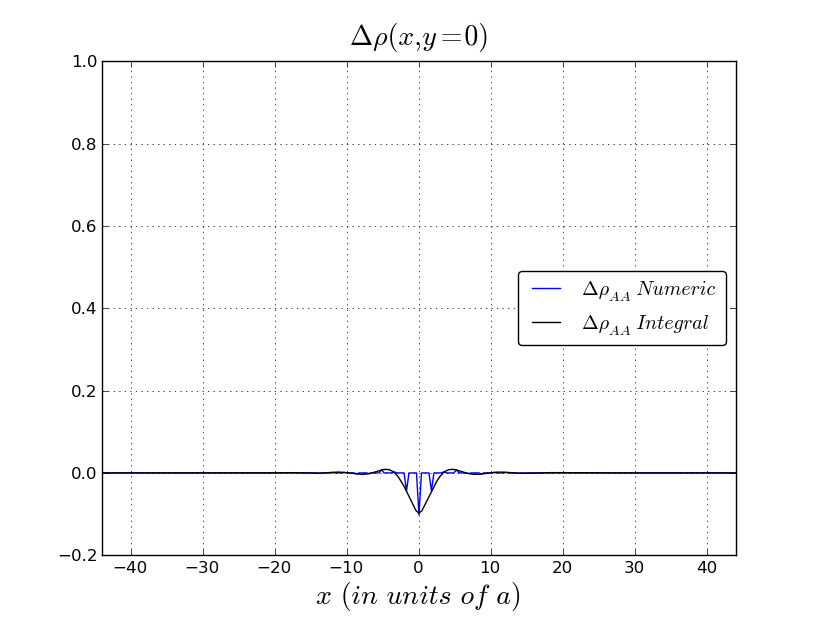}\\ 
\includegraphics[trim = 15mm 0mm 15mm 0mm, clip, width=5.5cm]{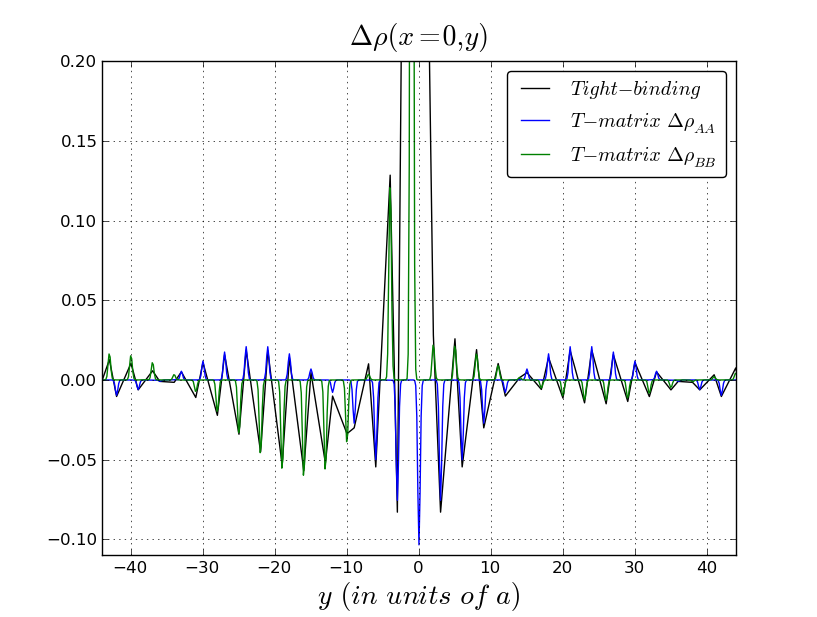}&
\includegraphics[trim = 15mm 0mm 15mm 0mm, clip, width=5.5cm]{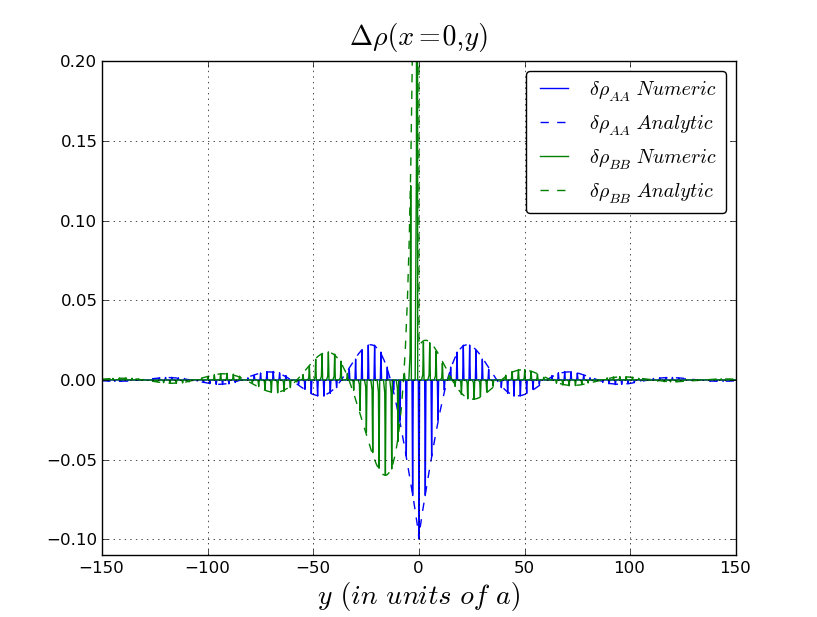}&
\includegraphics[trim = 15mm 0mm 15mm 0mm, clip, width=5.5cm]{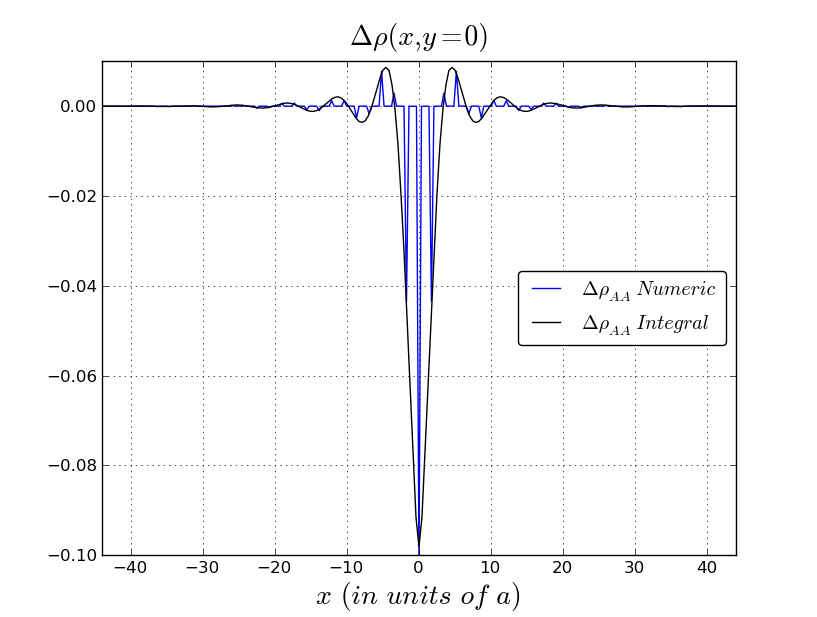}\\
\end{array}$
\caption{\small The LDOS correction as a function of position in the vicinity of the impurity. The second line presents a series of zooms-in of the plots outlined on the first line. In the first column we compare $\Delta \rho$ obtained using the full T-matrix approximation to the one obtained by the tight-binding method for an energy $\omega=-0.20t$. Note that, consistent to the low energy expansion, the FO obey an inverse square-root decay for large values of $y$ and are dephased by $\pi$ between the two sublattices. 
The second column presents a comparison between the correction to the LDOS $\Delta \rho$ along the $x=0$ direction obtained by the full T-matrix approximation (full lines) and by the low energy expansion (dotted lines) for $\omega=-0.20t$. In blue we plot the LDOS on the $A$ sublattice comprising the impurity ($y=0$) whereas in red the LDOS on the $B$ sublattice. 
In the third column we plot the LDOS along the $y=0$ direction obtained by the full T-matrix approximation for $\omega=-0.20t$. The blue curve is obtained using the full T-matrix approximation while the black one is obtained in the continuum approximation. }
\label{Analytic}
\end{figure*}

Now we turn back to the FO and evaluate the corrections to the LDOS using the corresponding expressions for the Green's function components in Eq.~(\ref{GreenFunctionExpression}). The results are presented in Fig.~\ref{Analytic}. We compare these results to a full evaluation of the T-matrix (without making the low-energy expansion), as well as with results obtained using the tight-binding method. As it can be seen in Fig.~\ref{Analytic} all methods yield very similar results, which confirms their accuracy for this type of calculation. We also note that the LDOS correction is asymmetric between the positive and negative values of $y$ on the $B$ sublattice, whereas it is symmetric on the $A$ sublattice. 

To obtain the asymptotic expansion of the Friedel oscillations we expand the Hankel functions for large values of $\omega y$ and we get: 
\begin{align}\label{LatticesLDOS}
&\Delta \rho_{AA}(0,y,\omega)\sim \frac{1}{y^{1/2}}\cos\Big(\frac{2\omega y}{c_y}+\pi\Big) \notag\\
&\Delta \rho_{BB}(0,y,\omega)\sim \frac{1}{y^{1/2}}\cos\Big(\frac{2\omega y}{c_y}\Big) 
\end{align} 
The resulting FO decay as $1/\sqrt{y}$ at large distances on both sublattices, slower than the typical inverse decay for a regular two-dimensional system, however their period is still proportional to $1/\omega$. When summing the contribution of the two sublattices, the terms in $y^{-1/2}$, which are dephased by a factor of $\pi$, vanish. The FO are then described by the next leading correction which is non-zero only on the $B$ sublattice: 
\begin{align}\label{LDOStot}
\Delta \rho(0,y,\omega)\sim \mp \frac{1}{\omega^{1/2}y}\cos\Big(\frac{2\omega y}{c_y}+\frac{\pi}{4}\Big).
\end{align} 
Here the minus/plus signs correspond to positive and respectively negative values of $y$. The long wavelength oscillations thus decay following the usual $1/y$ law, different from the $1/r^2$ law corresponding to the intra-nodal scattering in typical graphene. Thus the transition from the $1/r^2$ decay to a $1/r$ decay in the low-energy FO provides a real-space signature of the Dirac points merging. 

The Friedel oscillations along the perpendicular direction ($y=0$) cannot be evaluated analytically, however in the third column of Fig.~\ref{Analytic} we plot the dependence of the Friedel oscillations as a function of $x$ for $y=0$. Note that the amplitude of the oscillations is greatly reduced with respect of the oscillations in the $y$ direction, consistent with the asymmetric shape of the impurity-state cloud, elongated in the $y$-direction.

In Fig.~\ref{LDOS} we also present a two-dimensional plot of the LDOS at a finite energy, obtained both by using the full T-matrix form and the low-energy expansion. Note that the behavior is very similar to that obtained using the tight-binding method described in the previous section.


\section{Conclusion}
We have studied the LDOS in the presence of a simple impurity for an anisotropic graphene system at the Dirac-cone merging point. We have found that near this particular point, the zero-energy impurity wavefunction and the Friedel oscillations in the LDOS exhibit very peculiar features. In particular, the decay length of the Friedel oscillations along the anisotropy direction and along the direction perpendicular to this direction are very different, yielding a very asymmetric impurity state in real space.  The spatial dependence of the impurity state wavefunction allows us to clearly distinguish the semi-metallic phase (power-law decay of the wavefunction with the distance from the impurity) from the gapped phase (exponential decay). On the other hand, the semi-Dirac spectrum near the merging point induces a change of the decay laws in the Friedel oscillations from a inverse-square law ($1/r^2$) below the transition to an inverse-linear law ($1/r$) exactly at the transition. At low energy this provides a real-space signature of the topological transition. 

The agreement between the methods that we have used, the analytical T-matrix approximation, the numerical tight-binding exact diagonalization and the wavefunction considerations, is remarkable, proving the accuracy of these methods to describe the LDOS in the presence of disorder in a generic two-dimensional system.

\acknowledgments
We would like to thank J.-N. Fuchs, F. Piechon and G. Montambaux for interesting discussions and comments. This work has been supported by the ANR project NANOSIM GRAPHENE under Grant No. ANR-09-NANO-016, and by the FP7 ERC Starting Independent Researcher Grant NANO-GRAPHENE 256965.

\bibliographystyle{apsrev4-1}
\bibliography{references-1}


\end{document}